\documentclass[10pt,conference]{IEEEtran}
\IEEEoverridecommandlockouts
\usepackage[dvips]{graphicx}
\usepackage{color,array}
\usepackage{xspace}
\usepackage{amsmath}
\usepackage{caption}
\usepackage{subcaption}
\usepackage{placeins}
\usepackage{balance}
\usepackage{cite}
\usepackage{bm}
\usepackage{amsmath}
\usepackage{array}
\usepackage{setspace} 
\usepackage[nameinlink]{cleveref}
\usepackage{accents}
\usepackage{mathtools}
\usepackage{algpseudocode}
\usepackage[dvipsnames]{xcolor}
\usepackage{amsthm} 
\usepackage{multirow}
\usepackage[linesnumbered,ruled,vlined]{algorithm2e}
\usepackage{tabularx}
\SetKwRepeat{Do}{do}{while}
\usepackage{color}
\usepackage{epstopdf}
\usepackage{endnotes}
\usepackage{framed}
\usepackage{float}
\usepackage{cool} 
\usepackage{enumerate} 
\usepackage{pifont}
\Crefname{figure}{Fig.}{Figs.}

\usepackage{mathrsfs}
\usepackage[justification=centering]{caption}
\allowdisplaybreaks[4]

\newcolumntype{P}[1]{>{\centering\arraybackslash}p{#1}}

\DeclareUnicodeCharacter{2212}{-}

\usepackage{color}
\def\BibTeX{{\rm B\kern-.05em{\sc i\kern-.025em b}\kern-.08em
        T\kern-.1667em\lower.7ex\hbox{E}\kern-.125emX}}
\title{Empowering HWNs with Efficient Data Labeling: A Clustered Federated Semi-Supervised Learning Approach}
\author{\IEEEauthorblockN{Moqbel Hamood,  Abdullatif Albaseer, Mohamed Abdallah, Ala Al-Fuqaha}
\IEEEauthorblockA{ Division of Information and Computing Technology, College of Science and Engineering, \\ Hamad Bin Khalifa University, Doha, Qatar \\
\{moha19838, aalbaseer, moabdallah, aalfuqaha\}@hbku.edu.qa}
}
\begin{document}
\maketitle
\begin{abstract}
Clustered Federated Multi-task Learning (CFL) has gained considerable attention as an effective strategy for overcoming statistical challenges, particularly when dealing with non-independent and identically distributed (non-IID) data across multiple users. However, much of the existing research on CFL operates under the unrealistic premise that devices have access to accurate ground-truth labels. This assumption becomes especially problematic in, especially hierarchical wireless networks (HWNs), where edge networks contain a large amount of unlabeled data, resulting in slower convergence rates and increased processing times—particularly when dealing with two layers of model aggregation.
To address these issues, we introduce a novel framework—Clustered Federated Semi-Supervised Learning (CFSL), designed for more realistic HWN scenarios. Our approach leverages a best-performing specialized model algorithm, wherein each device is assigned a specialized model that is highly adept at generating accurate pseudo-labels for unlabeled data, even when the data stems from diverse environments.
We validate the efficacy of CFSL through extensive experiments, comparing it with existing methods highlighted in recent literature. Our numerical results demonstrate that CFSL significantly improves upon key metrics such as testing accuracy, labeling accuracy, and labeling latency under varying proportions of labeled and unlabeled data while also accommodating the non-IID nature of the data and the unique characteristics of wireless edge networks.
\end{abstract}
\section{Introduction}
 The proliferation of mobile devices and the Internet of Things (IoT) has not only exponentially increased data generation but also ignited a technological revolution that transforms everyday life~\cite{cao2021resource,lim2020federated}. This surge in data has enriched various applications and enabled cross-domain knowledge sharing, especially within Hierarchical Wireless Networks (HWNs) in which different edge networks, including mobile and IoT devices, are connected through a cloud~\cite{sun2019application,hosseinalipour2020federated}. Particularly, HWNs enable seamless collaboration among diverse IoT ecosystems, optimizing the data's utility for multiple applications by employing Machine Learning (ML) and Deep Learning (DL) techniques. However, traditional ML/DL approaches face considerable challenges when operating in HWNs, including handling vast volumes of data, which mostly is unlabeled, limited network resources, latency due to two levels of connections, and privacy concerns due to data leakage~\cite{feng2022mobility,vahidian2023rethinking,bertino2020artificial}.
 
In response, hierarchical federated learning (HFL)~\cite{luo2020hfel} was introduced as a proposed solution to address some of the aforementioned challenges. In the HFL approach, IoT-enabled devices conduct local model training and send only model parameters (i.e., weights and biases) to their corresponding edge servers and then to the cloud to update a global model. This strategy minimizes communication overhead and protects sensitive data. Despite its advantages, HFL approaches face several challenges during training. These include the diversity of device capabilities and the massive amount of non-independent and identically distributed (non-IID) data, which hinder efficient model development and communication. 

Cluster federated learning (CFL) paradigm~\cite{sattler2020clustered} has emerged as a promising solution that copes with these problems, especially the non-IID problem where the data residing on each device does not follow the same distribution, nor are the data points independent of each other. CFL is considered a post-processing technique that improves the FL algorithm without altering its fundamental communication protocol and at least achieves the same performance as FL. This is done by splitting devices into clusters based on similarity in their data distributions when the HFL objective at each edge network reaches a stationary point (i.e., where the FL model cannot improve further on its current data). Each cluster then develops a tailored model that accurately captures specific data patterns. Several existing research~\cite{sattler2020clustered,ghosh2020efficient,luo2021energy,gong2022adaptive} have delved into CFL's effectiveness within the edge network, focusing on fully labeled datasets to cluster the devices into different clusters. These works have also addressed bandwidth and energy consumption for the CFL in wireless edge networks. For example, the work in~\cite{albaseer2023fair} proposed solutions to prioritize participant devices according to their training latency by leveraging the dynamic characteristics of edge networks, scheduling devices based on their round latency, satisfying fairness, and exploiting bandwidth reuse. In addition, the work in~\cite{hamood2023clustered} proposed a solution to address the service heterogeneity where local models have different structures in HWNs using CFL with knowledge distillation.  However, none of these works \cite{sattler2020clustered,ghosh2020efficient,luo2021energy,gong2022adaptive,albaseer2023fair,hamood2023clustered} addressed the problem of unlabeled data under HWNs and CFL settings.

In the literature, applying  Semi-supervised learning (SSL) in conjunction with FL has received considerable effort~\cite{albaseer2022semi,diao2022semifl}.  For instance, the authors of~\cite{diao2022semifl} proposed a framework that integrates SSL and FL, allowing devices with only unlabeled data to train models across several local epochs, with the server maintaining a small subset of labeled data. Another study in~\cite{albaseer2022semi} introduced the federated semi-supervised learning (FedSemL) method, considering wireless communication challenges such as bandwidth, latency, and energy consumption.  Nevertheless, none of these works also considered the CFL and HWNs setting when dealing with the unlabeled data, as these methods generally focus on a single global model, which is insufficient for confidently labeling a vast array of non-IID data. 

In practice, the predominant data type collected by edge devices in HWNs is both unlabeled and non-IID, necessitating more robust solutions that account for limited edge network resources. Particularly, there is a need for an efficient approach to leverage the unlabeled data under CFL and HWN settings, in which CFL creates multiple specialized models only based on the fully labeled data. Each specialized model for each group of users or devices may fail to label the unlabeled data that belong to a different data distribution than the model belongs to. Another aspect is to consider the complexity of SSL when applied within HWNs due to dual-level model aggregation and computation. Therefore, exploring a framework that enables efficient use of unlabeled data in different edge networks while considering the unbalanced and non-IID data distribution is imperative.

Motivated by these gaps, we propose a novel framework incorporating CFL with  SSL in HWNs to bridge existing gaps in the research field. This framework, dubbed Clustered Federated Semi-Supervised Learning (CFSL), is designed to use the best-specialized models resulting from the clustering process in the CFL to label unlabeled non-IID data using the SSL. Rigorously, the aim is to accurately label the data samples using a specialized model, even if it has resulted from a different distribution, while optimizing the resources and adhering to deadline constraints at both edge and cloud servers. We formulate an optimization problem and introduce the proposed methods for a tractable solution. To assess the efficacy of the proposed approach, we conduct extensive simulations utilizing the FEMNIST dataset, evaluating testing accuracy, labeling accuracy, and latency compared to baselines.

The structure of this paper is as follows: Section \ref{Sys_model} details the system model, learning process, and computational and communication models. Section \ref{problem_form} delves into the problem formulation. The proposed approach is explained in Section \ref{proposed_sol}. Section \ref{results} presents our experimental findings and results. Finally, Section \ref{conclusion} concludes the paper and hints at potential future directions.
 \begin{figure}[t]
\centering
\includegraphics[width=0.75\linewidth]{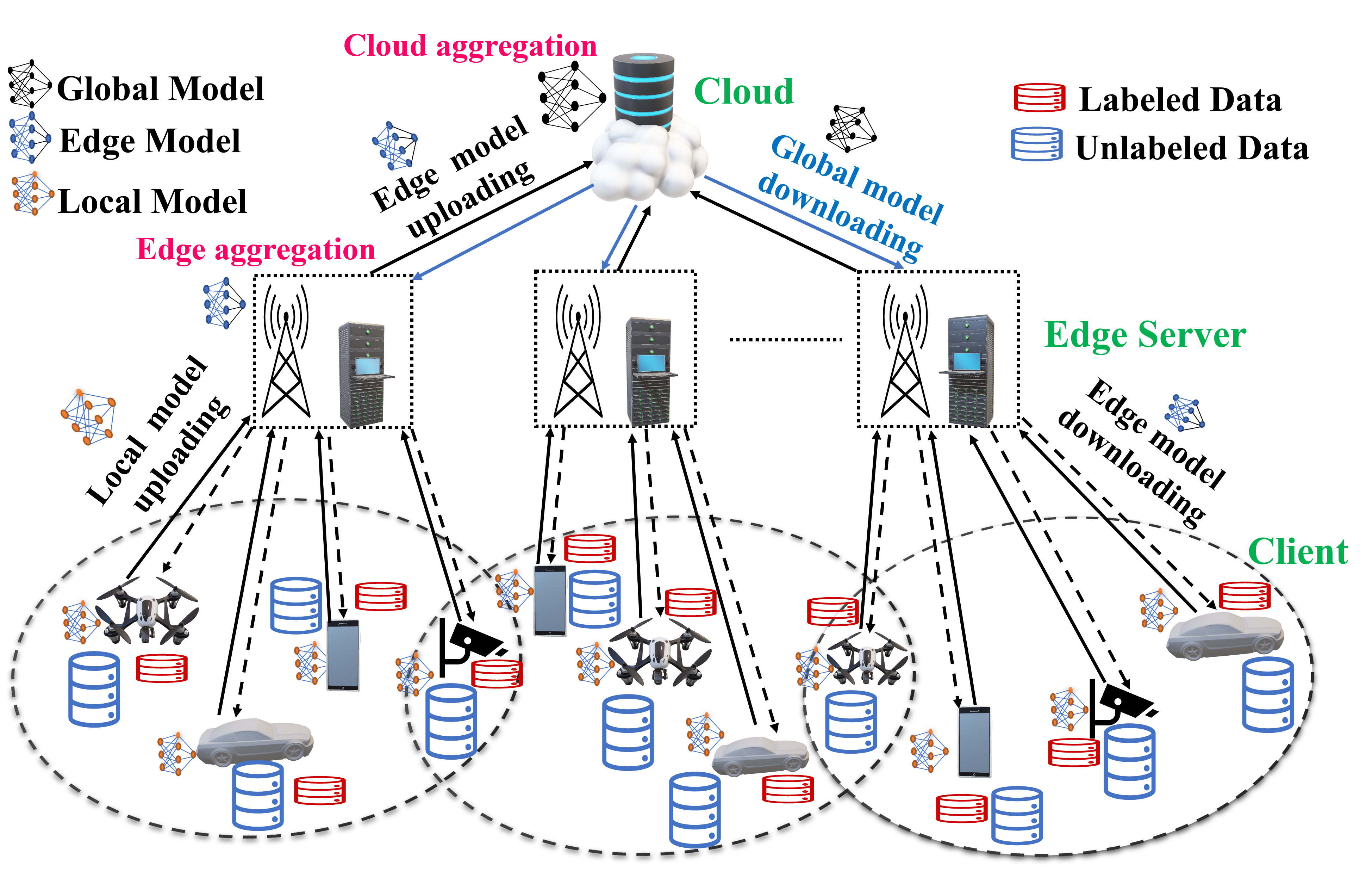}
\caption{{The system model}}
\label{fig:figure1}
\end{figure}
\section{System Model}
\label{Sys_model}
The system model illustrated in Fig. \ref{fig:figure1} comprises a cloud server connected to a set of edge servers $\mathcal{N}=\{1,\dots, N\}$, where $N=|\mathcal{N}|$, linked to a set of devices $\mathcal{K}=\{1,\dots, K\}$, where $K=|\mathcal{K}|$. Each $k$-\textit{th} device owns a local dataset $\mathcal{D}_k$, where a small set of this data is labeled, $\mathcal{D}^l_k=\{{x}^{(k)}_{i,d}\in \mathbb{R}^d,~ y^{(k)}_{i}\}_{i=1}^{D^l_k}$, and  much larger is unlabeled, $\mathcal{D}^u_k=\{{x}^{(k)}_{i,d}\in \mathbb{R}^d\}_{i=1}^{D^u_k}$. Here, ${x}^{(k)}_{i,d}$ is the $d$ dimension feature vector of the $i$-\textit{th} sample at $k$-\textit{th} device, and $y^{(k)}_{i}$ is the corresponding label; thus, the entire dataset $\mathcal{D}_k=\mathcal{D}^l_k\cup\mathcal{D}^u_k$. The total number of  labeled and unlabeled data  for the $k$-\textit{th} device is defined as ${D}^l_k=|\mathcal{D}^l_k|$ and ${D}^u_k=|\mathcal{D}^u_k|$, respectively.
It is worth noting that devices are associated with edge servers based on physical proximity.
\subsection{HFL Model}
In HFL, we assume the cloud server connects different edge networks, each with an edge server and a set of connected devices. To start the HFL process, the cloud initiates the global model, $W_\circ$, and sends it to the edge servers-connected devices. Each edge server selects a subset of active devices ($S^n_r$) to participate in the training. Each device trains its local model using the labeled data and uploads it to the associated edge server. Afterward, the edge servers aggregate and average local models from their devices before sending them to the cloud. The cloud receives the models, updates the global model, and then forwards it back to the edge servers and the corresponding devices for further training. This process iteratively continues until the global model converges. It is worth mentioning that the cloud and edge servers enforce a deadline constraint to synchronize model updates to attenuate the risk of latency-induced delays resulting from the time-varying channels in wireless networks. Let us assume that stochastic gradient descent (SGD) is used for training with mini-batches as a local solver to update the local models using their labeled data. The batch in mini-batch SGD performs several updates within one epoch defined as $\sigma=\frac{{D}^l_k}{b}$, where $b$ is the batch size. 

To evaluate the performance of local models, each  $k$-\textit{th} device runs the min-batch SGD algorithm $\sigma$ times to minimize the error of the local model between the actual and predicted labels in every $r$-\textit{th} round.  Thus, the local loss function for the $k$-\textit{th} devices to catch the error over each $i$-\textit{th} sample of labeled data $\mathcal{D}^l_k$  is defined as:
\begin{equation}
    F_k(W_r):=\frac{1}{D^l_k} \sum_{i \in \mathcal{D}^l_k} f_i(W_r).
\label{loss-func}
\end{equation}
The SGD iterates through each sample point to reduce its loss function (i.e.,  $f_i(W_r)= \mathcal{F}_i(x^{(k)}_i,y^{(k)}_i; W_r)$), and achieve an optimal local model that fits well with its labeled data distribution.
Ultimately,  HFL aims to minimize the total loss function over all devices' labeled datasets $\mathcal{D}^l = \bigcup_{k} \mathcal{D}^l_k$.
\subsection{Clustered Federated Learning (CFL)}
The CFL process in HWNs can be represented as parameterized trees, where the roots of these trees represent the HFL processes that reach stationary solutions (i.e., $W^*_n$). After that, the CFL partitions the devices in each $n$-\textit{th} edge network into two clusters. This occurs by requesting the devices' gradients and then computing their similarity to split them into clusters, each with a specialized model.  In practice, the cosine similarity ($\alpha$) is used by the edge server to compute the similarity between any two devices' gradients within the same  $n$-\textit{th} edge network or by the cloud server to compute the similarity between different edge networks in HWNs. Formally, the cosine similarity in this context is obtained as follows:

\begin{footnotesize}
\begin{align}
\alpha^n_{k,k'}&:=\alpha(\nabla F_k(W^*_n), \nabla F_{k'}(W^*_n)):=\frac{\langle  \nabla F_k(W^*_n),\nabla F_{k'}(W^*_n)\rangle}{\parallel\nabla F_k(W^*_n) \parallel \parallel\nabla F_{k'}(W^*_n)\parallel} 
\nonumber \\ &= 
\begin{cases}
1, & \text{if $J(k)=J(k')$}\\
-1, & \text{ if $J(k)\neq J(k')$},
\end{cases}
    \label{similarityformula}
\end{align} 
\end{footnotesize} 
Here, $J(k)$ and $J(k')$ are the distributions of local labeled data for $k$ and $k'$ devices, respectively. Consequently, the accurate bi-partitioning for clusters can be formulated as $c_{n_1}=\{k| \alpha^k_{k,0}=1\}, \quad c_{n_2}=\{k| \alpha^n_{k,0}=-1\}$. To perform accurate splitting between two clusters, two critical conditions must be satisfied. First, the solution of the HFL objective must approach the stationary points, mathematically expressed as follows: $ \tiny
  0  \leq \Biggl\|\sum_{k\in \mathcal{K}_n}\frac{D^l_k}{|D|} \nabla_W F_k(W^*_n)\Biggr\| < \varepsilon_1,\label{cond1}$
 and second,  devices must be far from the stationary point of their respective local loss functions: $\tiny \max_{k \in \mathcal{K}_n} \left\| \nabla_W F_k(W^*_n)\right\|>\varepsilon_2>0,
\label{cond2}$
where $\varepsilon_1$ and $\varepsilon_2$ are pre-determined hyperparameters designated explicitly for controlling the clustering task using the labeled datasets.
It is worth highlighting that training the models using labeled data continues in each cluster until either it approaches a new stationary point (i.e., splitting is required) or reaches the stopping point (i.e., no further splitting for this model). This process recursively continues until all clusters reach the stopping point.
\subsection{Communication and Computation Models}
\noindent\textbf{Computation model:} Let us define $f_k$ as the computation frequency (CPU speed) and $\Theta$ as the needed CPU cycles to process one data sample. Therefore, the local computation time for $k$-\textit{th} device running the local solver for $P$ epochs in a given round $r$ is obtained as follows: $T^{cmp}_k=P\frac{{D}^l_k\Theta}{f_k}$.
In the edge servers layer, we have to set a deadline $T^{n}_{S_r}$ in every $r$-\textit{th} round to avoid longer waiting times for devices that take a longer time than others to complete their assigned tasks during the training process due to large data sizes, limited resources, or intermittent connectivity.\\
\textbf{Communication model:} First, the uploading time of models to the edge server is defined as $T^{com}_k=\frac{\delta}{R^{com}_k}$, where $\delta$ is the model size and ${R^{com}_k}$ is the transmission data rate achieved by the $k$-\textit{th} device, defined as ${R^{com}_k = \beta_{k,n}~B_n~\text{log}_2\left(1 + \frac{| \mathbf{h}_k|^2 P_k}{ N\circ }\right)}$. Here, $\beta^r_{k,n}$ indicates the bandwidth allocation ratio, $\beta^r_{k,n} B_n$ is the allocated bandwidth for the $k$-\textit{th} device, $h_{n,k}$ is the channel gain between the $k$-\textit{th} device and the $n$-\textit{th} edge server, $ N_\circ$ refers to the complex additive white Gaussian noise (AWGN) power, and $P_k$ is the transmission power. In addition,  we use orthogonal frequency-division multiple access (OFDMA) techniques to partition the channel into sub-channels. Each edge server has limited sub-channels to upload local models in every $r$-\textit{th} round.
Accordingly, the total time required for each $k$-\textit{th} device for training and uploading to the $n$-\textit{th} edge server is  defined as:
\begin{equation}
    T_{S_r}^{n}=\mathrm{max} \left( T^{cmp}_k+ T^{com}_k\right).
\end{equation}
It is important to note that, given the edge servers' robust processing and transmission capacities, we neglected the time and energy associated with broadcasting edge and specialized models to devices.

For the cloud layer, the duration taken by the \(n\)-\textit{th} edge server to upload its models can be defined as \(T^{\mathrm{cloud}}_{n} = \frac{\delta}{R^{com}_{n}}\), where \(R^{com}_n\) signifies the transmission rate achievable by the \(n\)-th edge server. 
Accordingly, the latency in a given round \(r\)-\textit{th} global round is expressed as follows:
\begin{equation}
    T_r=\underset{n\in \mathcal{N}} \max (T^{\mathrm{cloud}}_{n}+T_{S_r}^{n}).
\end{equation}
\section{Problem Formulation}
\label{problem_form}
As previously mentioned, each device holds a substantial pool of unlabeled data. Importantly, each device is allocated a specialized model based on its fully labeled data. However, this specialized model may not necessarily be suitable for labeling the device's unlabeled data samples, which could belong to a different data distribution requiring another specialized model or refining the whole training process. This adds another layer of complexity to the challenge, particularly when dealing with non-IID data distributions. Our objective is to harness these unlabeled samples effectively to enhance overall system performance. Specifically, selecting the best-performing specialized model for each device, potentially spanning different edge networks, is critical. This task becomes even more daunting when considering the limited-resource aspects of network configurations, such as computational power and communication bandwidth. When incorporating unlabeled data predicted with a probability exceeding a given threshold, $\phi$,  meticulous design is imperative for the CFSL approach. Mathematically speaking, we can express the pseudo-labeled data $D^l_{semi}$ that satisfy the confidence threshold $\phi$ as follows:
\begin{equation}
D^{l,new}_{k}= D^l_k+D^l_{semi}.
\label{new_dataset}
\end{equation}
From  (\ref{new_dataset}), it is crucial for the $k$-\textit{th} device that adds new samples to its training data to adjust $T^{cmp}_k$ to meet the deadline constraint as $T^{cmp, new}_k=P\frac{{D}^{l, new}_{k}\Theta}{f_k}.$
    
To better refine our objective, we aim to minimize the loss function $F(W_k,\mathcal{D}^l_k )$ for each device's labeled data while maximizing the utility function $\mathcal{U}(W_m,\mathcal{D}^u_k )$  of using model $m$ to label the unlabeled data. This occurs by firstly deriving an optimal aggregation methodology to attain $M$ specialized models $\mathcal{M} = $\(\{W_m: m = 1, \dots, M\}\) within a total time budget, \(T_{tot}\), and secondly selecting a specialized model that provides optimal pseudo labeling for this device. In this regard,  the selected sets of available and active devices for all rounds ($R$) are denoted as $S_{[R]}=[S_1,\dots,S_R]$, and the deadline for every round is $T_{[R]}=[T_1,\dots,T_R]$. Therefore, the aim is to minimize the loss of all specialized models while maximizing the utility of using the unlabeled data, which can be defined as:
\begin{equation}
\small
 f(W, z, \mathcal{D}^l, \mathcal{D}^u)=\sum^{R}_{r=1}\sum^{K}_{k} \left[ F(W_k,\mathcal{D}^l_k ) - \lambda \sum_{m \in \mathcal{M}} z_{km} \mathcal{U}(W_m, \mathcal{D}^u_k) \right] 
\end{equation}
Consequently, the whole problem formulation for all clusters can be written as follows: 
\begin{subequations}
\small
\begin{align}
\textbf{P1:} \quad \underset{\mathbf{W},R, \mathbf{z}}{\text{minimize}}
&~f(W, z, \mathcal{D}^l, \mathcal{D}^u) \label{eq:OptmizedProblem1}\\
\textrm{s.t.} \quad  
& F(W_{m}) - F(W^*_{m}) \le \epsilon, \quad \forall m \in \mathcal{M}, \label{eq:convergence_constraint} \\
& \sum_{r=1}^{R} T_r(S_r)\leq T_{tot}, \quad r\in \forall [R], \label{total_time}\\
&(T^{cmp,new}_k+T^{com}_{k}) \leq  T_{S^n_r}^{n},~\forall k \in \mathcal{K},~\forall n \in \mathcal{N} \label{Deadline_one_client}\\
& \sum_{k \in \mathcal{K}} \text{B}(W_k) \leq B, \label{bw_constraint}\\
& \sum_{m \in \mathcal{M}} z_{km} = 1,  \quad \forall k, \in \mathcal{K},
\label{single_model_selection}\\
&  z_{km} \in \{0,1\}, \quad \forall k\in\mathcal{K},\quad m \in \mathcal{M}
\label{model_indicator}\\
& A_k \in \{0,1\},\quad \forall k \in \mathcal{K},
\label{association}
\end{align}
\end{subequations}
where $\lambda$ refers to a trade-off parameter. Constraint (\ref{eq:convergence_constraint}) ensures that each cluster's specialized models converge over the labeled and pseudo-labeled data. Constraint (\ref{total_time}) ensures that the total time taken by all selected devices in all edge networks during the CFL process remains within the predefined time allocation, $T_{tot}$ across all rounds. Constraint (\ref{Deadline_one_client}) ensures that the combined computation and communication time in a given round does not surpass the time limit set by each edge server. Constraint~(\ref{bw_constraint}) guarantees that the total bandwidth used to transmit the models across all devices should not exceed the system bandwidth ($B$). To ensure that each $k$-\textit{th } device uses only one specialized model $W_m$ (it could belong to the device's cluster or from other clusters) for the pseudo-labeling process.  Constraint~(\ref{single_model_selection}) is a binary variable indicating whether the model $m$ is used to label the unlabeled data of device $k$ the model ( $z_{km}=1$) if yes,  ($z_{km}=0$ otherwise). The final constraint~(\ref{association}) is a binary association constraint for $k$-\textit{th} device (i.e., $1$ if the device is allowed to connect to a given edge server, $0$ otherwise).
\section{Proposed Solution}
\label{proposed_sol} 
As seen from \Cref{problem_form}, Problem \textbf{P1} is an NP-hard and MIP problem, primarily because the product of the binary variable $z_{km}$ with the utility function results in a non-linear relation. Also, the problem contains the continuous variable $W_m$ and the integer variable such as $z_{km}$. In addition, it requires determining the impact of $W$, $R$, and $z$ on each cluster model's weight vector, $F(W_m)$. Further complicating the problem are the unpredictable variations in wireless channels and inconsistent computation times across devices during training. All these complexities complicate selecting the best-performing specialized model for pseudo-labeling, preventing a direct solution.   
Thus, this section introduces a tractable proposed solution to ensure proper labeling under HWNs and CFL settings while considering the network limitations. 
However, the critical aspect of this approach is how to select the best-performing specialized model and how to determine the appropriate time to deploy such a model for predicting and injecting the unlabeled data. In CFL, a device is allocated only a specialized model when the clustering starts. This process is imperative since the global model under the non-IID data distribution setting will never converge and accurately label the unlabeled data. For clarity, suppose a specialized model $m$ is assigned to a cluster and its devices. Evaluating whether this specialized model aligns with the unlabeled data distribution within the same cluster is crucial. If differences arise, we select the most suitable specialized model from other clusters within the same edge network or from different edge networks, which aligns with the current unlabeled data distribution and meets a predefined confidence threshold $\max(P(y^u_i)) \geq \phi$, where $\phi$ is the output probability, and $y^u_i$ is pseudo labels for the $i$-\textit{th} sample.  

Based on this, our proposed approach determines when and how to select the best-performing specialized model for the $k$-\textit{th} device to label the unlabeled data, which is outlined as follows. Initially, the cloud initiates $W_\circ$ and sends it to the edge servers, which then forwards it to the edge devices for training. Each device uses its labeled data to train the local model and upload it to the associated edge server. Afterward, the edge servers aggregate these local models, fuse them, and then send them to the cloud. The cloud also aggregates edge models, averages them, and sends them back to the edge servers and devices for further training. This process is repeated until the splitting conditions are satisfied. Specifically, each edge server splits the assigned devices into clusters, each with a specialized model that fits the data distribution well. Once we have the specialized models, each $k$-\textit{th} device starts predicting and injecting unlabeled data using the best-performing specialized model. It is worth noting that the best-performing specialized model for the $k$-\textit{th} device does not necessarily mean the specialized model where the device belongs during the first training phase; it could come from a different cluster in the same or a different edge network. This is because this specialized model may fail to accurately label the unlabeled data since the unlabeled data samples might have different data distributions. Therefore, the $k$-\textit{th} device will notify the cloud to send all the specialized models from different clusters.  After that, it will evaluate the utility of using every specialized model to label its unlabeled data.  Accordingly, the $k$-\textit{th} device selects only the best-performing specialized model with the highest labeling accuracy and less latency to label its unlabeled data perfectly.
Algorithm~\ref{CFL1} details the procedure steps of our proposed approach.
\begin{algorithm}
\scriptsize
\caption{CFSL Framework}
\label{CFL1}
\KwIn{Device count ($K$), initial parameters ($W_\circ$), controlling parameters $\varepsilon_1$ and $\varepsilon_2$, epoch count ($P$), batch size $b$, confidence level $\phi$, and device's labeled data (${D}^l_k$) and unlabeled data (${D}^u_k$)}
\KwOut{$M$ specialized models and a FL global model}
  
\textbf{Initialization:} Define device cluster $\mathcal{K}$, set $W_k \leftarrow 0~\forall k$, and $r=1$

\For{$r=1$ to $R$}{
  
  \textbf{Aggregation at Edge} \par
  
  \For{$n=1$ to $N$}{
      \textbf{Edge server $n$ pre-process:}
      Identify responsive devices and distribute $W_{r-1}$
      
      \textbf{Device tasks:}
      \For{$k=1$ to $|\mathcal{S}^n_{r}|$}{
          Get $W_{r-1}$.\\
           // \textbf{Device's Pseudo Labeling}\\
          Perform training: $W_k = W_{r-1} - \eta\sum_{t=1}^{\tau}\nabla F_k(W_k{(t)})$\\
          //\textbf{ Starting clustering, developing  specialized models}\\
          Label ${D}^u_k$ using the best-performing specialized model $m$\\
         Satisfying confidence threshold: $\max(P(y^u_i)) \geq \phi$\\
           Injecting pseudo-labeled data into the training data
      }
      Aggregate device models, compute $F_n(W_r)$, and adjust $W_{n,r}$\\
      \For{$c \in \mathcal{M}$}{
          Determine $\triangle W_m$ and update the specialized model set if splitting conditions are met
      }
      Update $\mathcal{M}$
  }
  
  \textbf{Cloud Aggregation:}\\
  {{
      Aggregate edge  and specialized models, and update global model $W_r$\\\If{$\textbf{M}>2$}{perform similarity checks among specialized models}  
  }}
  Increment $r$
}
Return global model and $M$ specialized models
\end{algorithm}
\section{Performance Evaluation}
\label{results}
\subsection{Experimental Setup}
This section evaluates the CFSL performance via simulation and elaborates on the experimental setup. It showcases the numerical results of the proposed approach in contrast to baseline techniques,  focusing on testing accuracy, labeling accuracy, and latency.\\
\textbf{Experimental Parameters:} The total bandwidth of the system is 10 MHz. Each device's channel gain ($\mathbf{h}_k$)  is determined based on a path loss (($\alpha=g\circ (\frac{d_\circ}{d})^4$)), where the reference gain $g_\circ$ is -$35$ dB and the reference distance $d_\circ$ is set at $2$ m. The AWGN power is maintained at $N_\circ=10^{-6}$ watts. The computation frequency  $f_k$ ranges from $1$ GHz to $9$ GHz. Every device requires a $\Theta=20$ cycle/sample. The transmission power, $P_k$,  confined within a range of  $P^{\min}_k=-10$ dBm and $P^{\max}_k=20$ dBm.
The simulation parameters are set for $200$ communication rounds (R) involving $200$ devices (K), while the learning rate is $0.01$ and we use $5$ epochs for local training in each global round.\\
\textbf{Dataset Preprocessing:} 
 We use the FEMNIST dataset~\cite{caldas2018leaf}, which has 62 classes (10 digits, 26 lowercase letters, and 26 uppercase letters). It is used under non-IID settings where each device is assigned a maximum of $2$ classes. This ensures that each edge network receives a distinct data subset, preventing any single device from having a complete view of the entire dataset.  We employed convolutional neural networks (CNN) for the classification task. We use only a small portion of labeled data samples and numerous unlabeled samples to prepare the data for our experiments.\\
\textbf{Performance Metrics and Benchmarks:} We evaluated the performance of our proposed approach against two different techniques. The first baseline is the CFL technique, which utilizes the fully labeled dataset for training. The second is the traditional HFL in conjunction with the SSL technique, which uses the global model for pseudo-labeling tasks.
\subsection{Results and Discussion}
 \begin{figure}[t]
    \centering
    \includegraphics[width=0.53\linewidth]{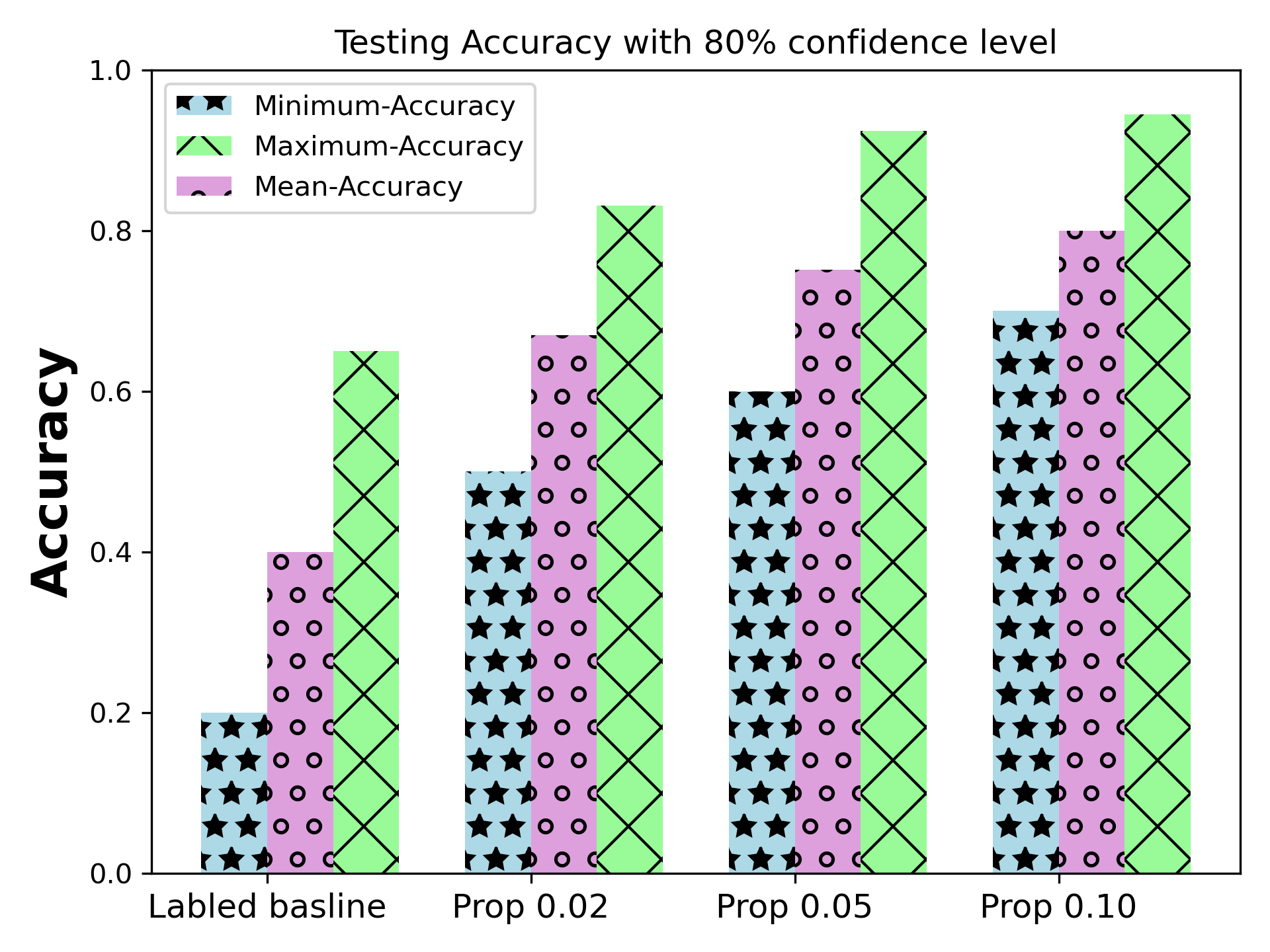}
    \caption{Testing accuracy for the proposed and baseline ( fully labeled CFL) when $\phi=0.8$.}
    \label{Label_VS_semi}
\end{figure}
 \begin{figure}[t]
    \centering
   \begin{subfigure}[b]{0.24\textwidth}
    \includegraphics[width=1\linewidth]{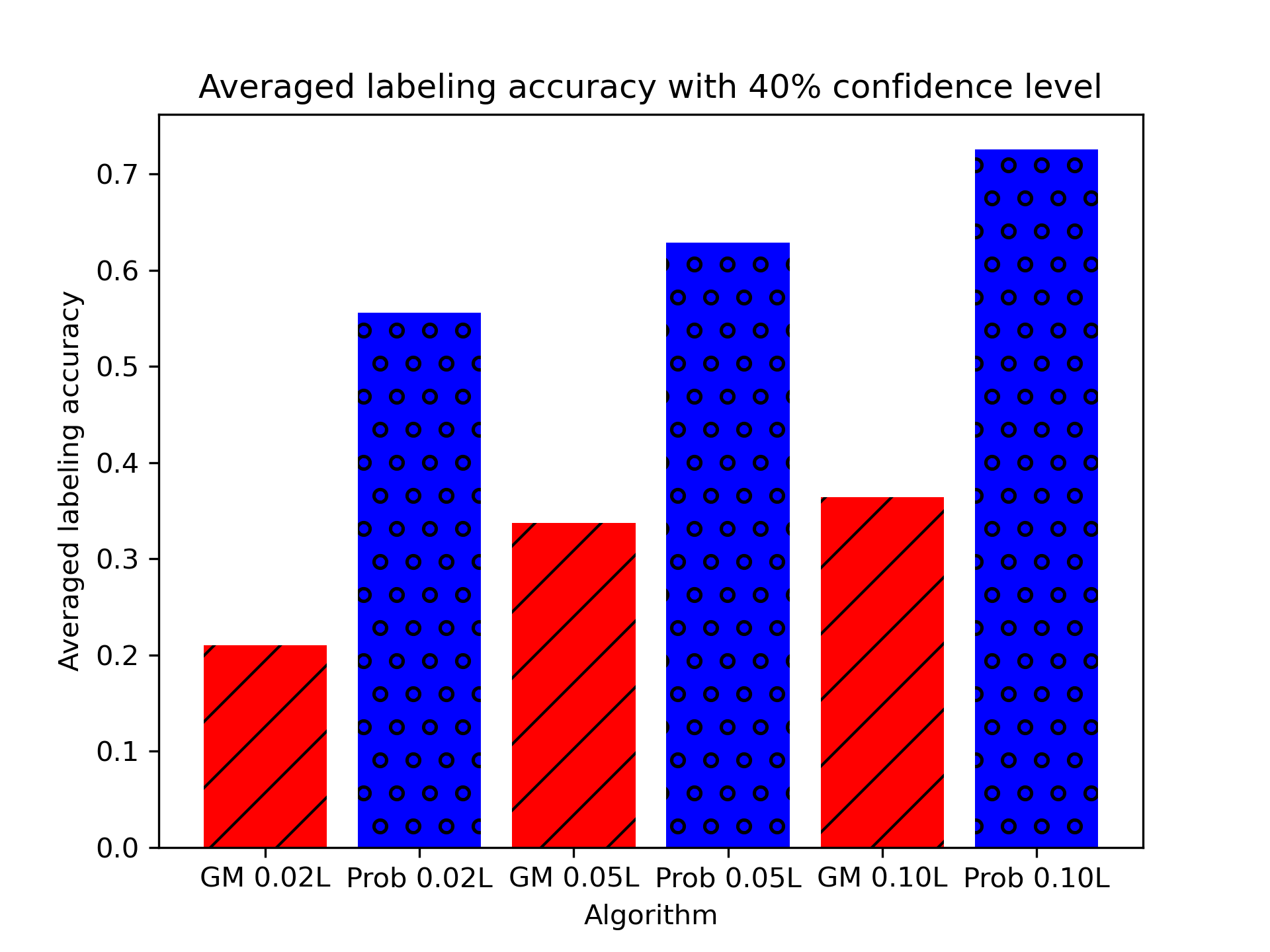}
    \caption{$\phi=0.40$}
    \label{acc1}
 \end{subfigure}
 \begin{subfigure} [b]{0.24\textwidth}
    \includegraphics[width=1\linewidth]{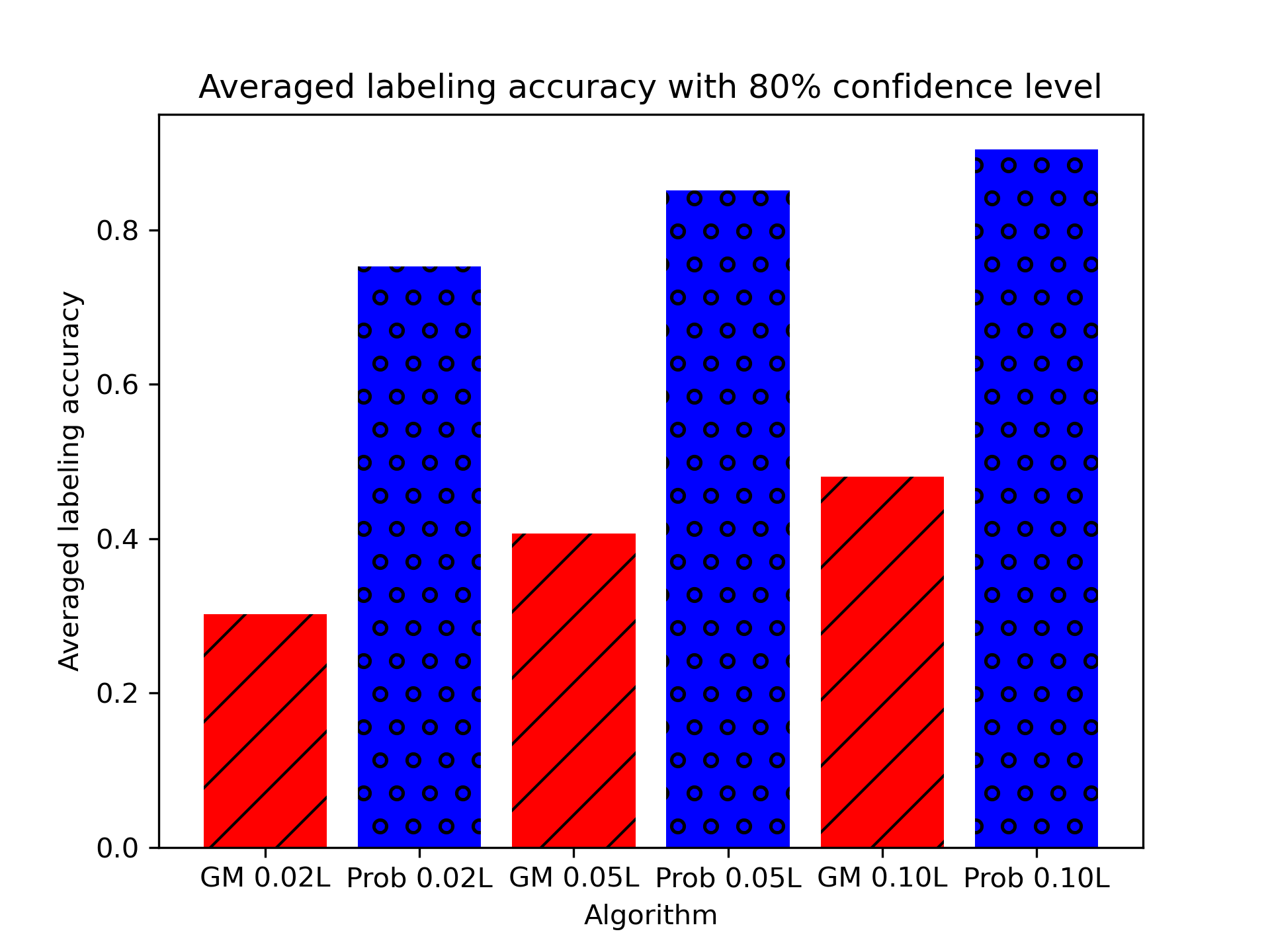}
    \caption{$\phi=0.80$}
    \label{acc2}
 \end{subfigure}
 \caption{Labeling accuracy for the proposed and baseline approaches with different percentages of labeled data.}
\end{figure}
 \subsubsection{The Significance of Applying CFSL}
 As shown in Fig. \ref{Label_VS_semi}, the testing accuracy is simulated using all participating devices' minimum, mean, and maximum accuracy to show how the proposed CFSL improves the performance. One can note that our proposed approach surpasses baselines and significantly improves the testing accuracy at different percentages of labeled data (i.e., $2$$\%$, $5$$\%$, and $10$$\%$). This stems from injecting unlabeled data into the training data during the training phase using the most fitting model for the local unlabeled data, enhancing the overall system performance.
 \subsubsection{Labeling Accuracy and Latency}
 Figs. \ref{acc1} and \ref{acc2} illustrate the labeling accuracy of the proposed and baseline algorithms when the confidence level is $\phi=0.40$ and $0.80$, respectively. Notably, injecting the unlabeled samples using varying proportions of labeled data and confidence levels through the specialized models significantly enhances labeling accuracy, outperforming baselines. One can notice that the increment of labeling accuracy in CFSL is much larger than the baseline using different percentages of labeled data (i.e., 0.02, 0.05, and 0.10). This stems from the fact that specialized models adapt more precisely to the unique features of the data with a faster convergence rate and perform better prediction of labels in non-IID data, even if unlabeled data samples belong to different distributions. In contrast, the global model fails to capture all data patterns due to the non-IIDness of the data, leading to potential inaccuracies and slowing down convergence rates during incorrect labeling or the delay of injecting the unlabeled samples.  
 \begin{figure}[t]
  \centering
 \begin{subfigure}[b]{0.24\textwidth}
    \includegraphics[width=1\linewidth]{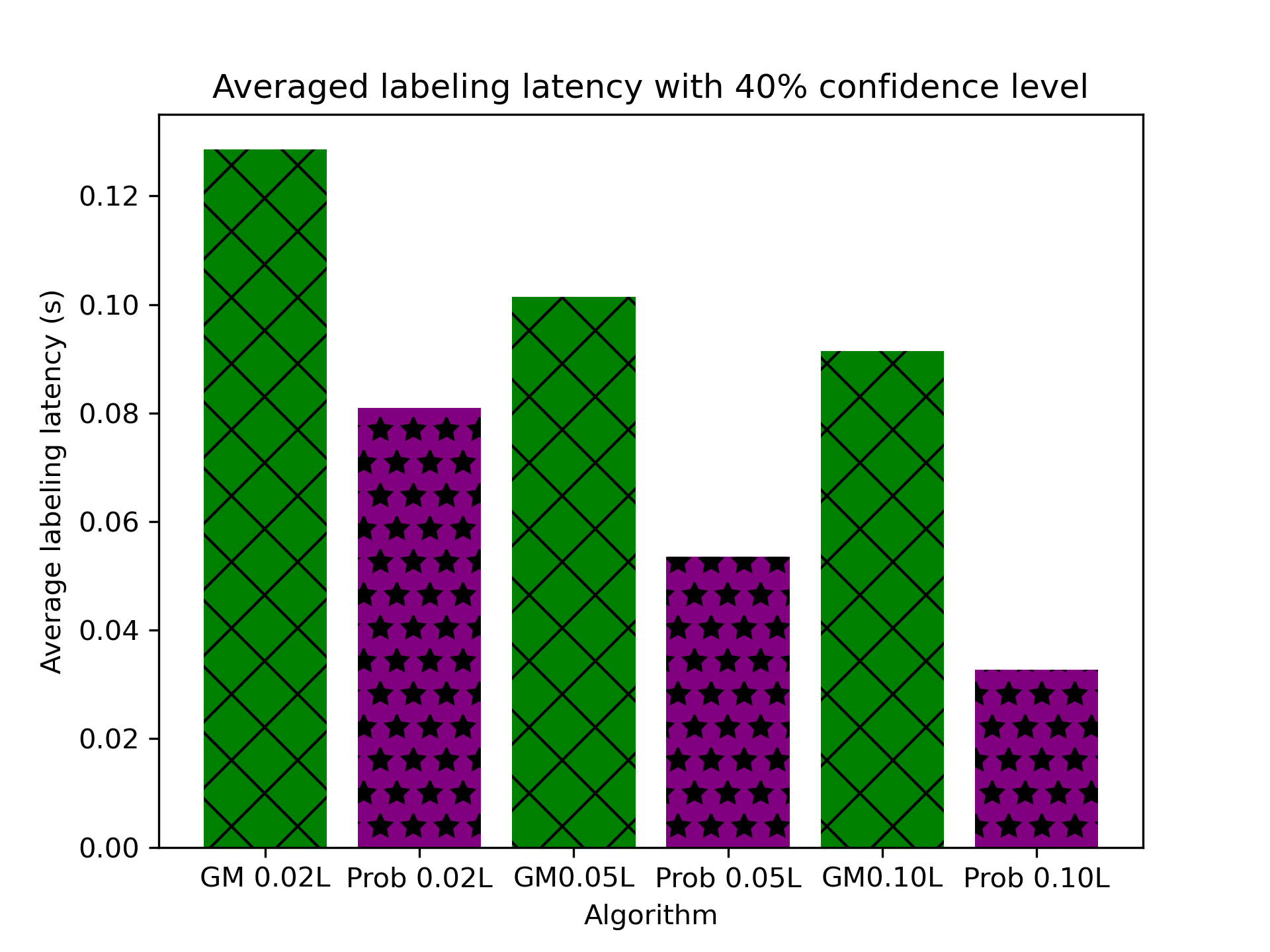}
    \caption{ $\phi=0.40$}
    \label{acc3}
\end{subfigure}
\begin{subfigure}[b]{0.24\textwidth}
    \includegraphics[width=1\linewidth]{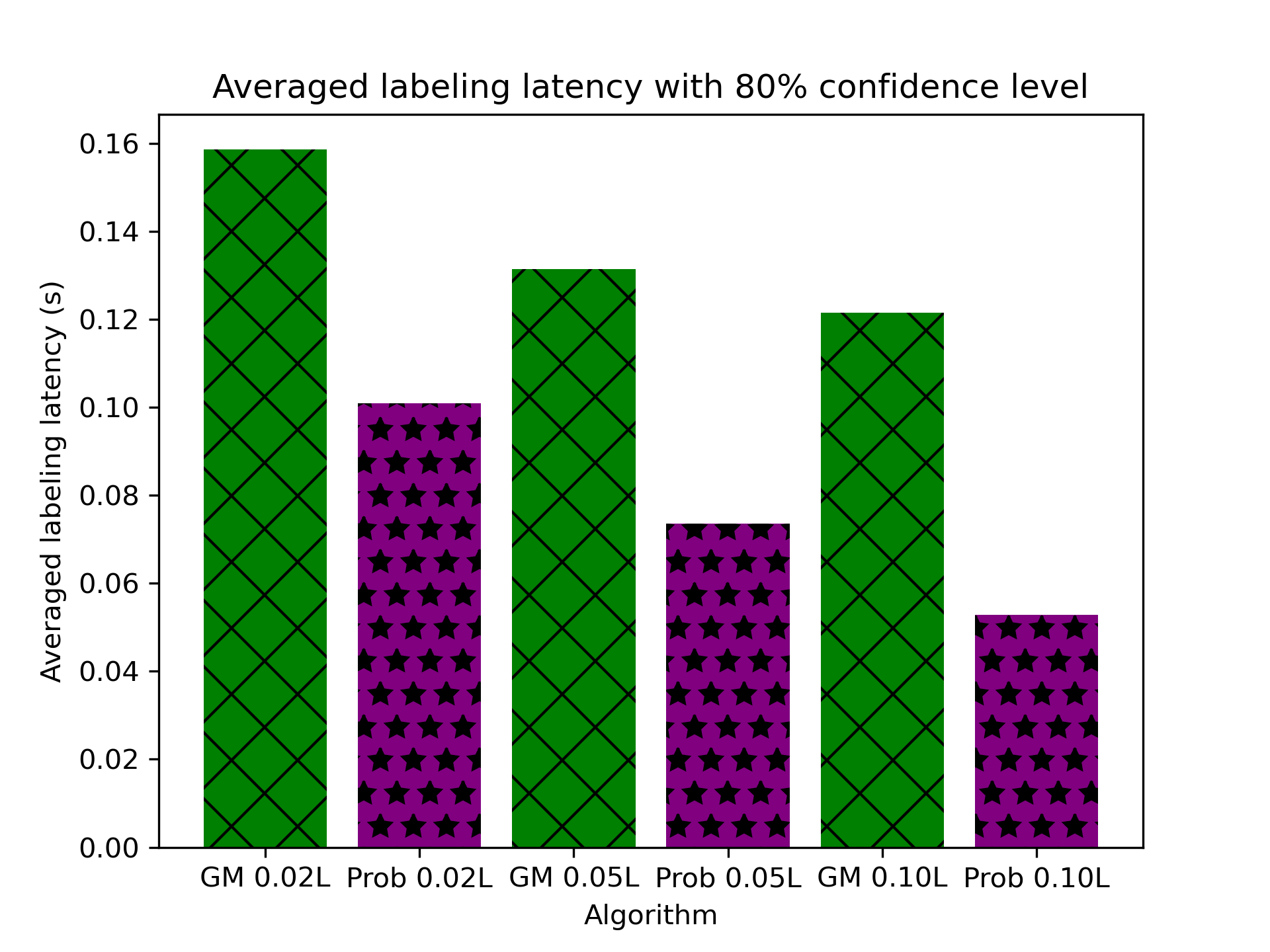}
    \caption{ $\phi=0.80$}
    \label{acc4}
\end{subfigure}
\caption{Labeling latency for the proposed and baseline approaches with different percentages of labeled data}
    \label{fig:enter-label2}
\end{figure}

Regarding the labeling latency, Figs. \ref{acc3} and \ref{acc4} showcase the averaged labeling latencies of both the proposed approach and baselines at confidence levels of 
 $\phi=0.4$  and $\phi=0.8$. CFSL consistently exhibits a much lower latency than the baselines across different labeled sample proportions, even when only $2\%$ of labeled data is used. This efficiency is attributed to selecting the most fitting specialized models, which are fine-tuned for distinct tasks, allowing for faster processing and minimizing computational demands compared to using only a single global model as in baselines. Furthermore, elevating the confidence level from  $\phi=0.4$  to $\phi=0.8$ results in a substantial rise in latency, highlighting the trade-off between computational demands and confidence-level accuracy.
\section{Conclusion}
\label{conclusion}
In this paper, we addressed the significant challenge of enabling efficient CFL and SSL in HWNs under the constraints of non-IID data distribution and limited network resources. We formulated an optimization problem to achieve our goal. Due to the NP-hard and MIP nature of the problem, finding an optimal solution is computationally intensive and not practical; thus, we proposed a tractable solution to determine how and when to use the best-performing specialized model for labeling, balancing trade-offs between accuracy, latency, and system constraints. Our approach started by initializing a global model that is iteratively refined based on edge devices' local models. CFL recursively grouped devices into clusters, and each attained a specialized model that evaluated its utility in labeling the unlabeled data of devices within its cluster and potentially across clusters. The specialized model offering the highest accuracy and lowest latency in labeling a device's unlabeled data is selected for that task. Our approach also effectively balanced the trade-off between labeling accuracy and system constraints, offering a scalable and efficient solution for harnessing unlabeled data in HWN and CFL settings.
\section*{Acknowledgement}
This publication was made possible by NPRP-Standard (NPRP-S) Thirteen (13th) Cycle grant \# NPRP13S-0201-200219 from the Qatar National Research Fund (a member of Qatar Foundation). The findings herein reflect the work and are solely the authors' responsibility.
\bibliographystyle{IEEEtran}
\bibliography{Ref}
\end{document}